# A Hybrid Approach: Utilising K-means Clustering and Naive Bayes for IoT Anomaly Detection

Lincoln Best[1], Ernest Foo [1] [0000-0002-3971-6415] and Hui Tian [1] [0000-0002-7952-571X]

[1] School of Information and Communication Technology, Griffith University, Brisbane, QLD, 4111, Australia
Lincoln.best@griffith.edu.au
Ernest.foo@griffith.edu.au
Hui.tian@griffith.edu.au

**Abstract.** The proliferation and variety of Internet of Things (IoT) devices means that they have increasingly become a viable target for malicious users. This has created a need for anomaly detection algorithms that can work across multiple devices. The suggested algorithm will further ensure that our data remains secure from malicious users thus potentially avoiding related real-world issues. This thesis suggests a potential alternative anomaly detection algorithm to be implemented within IoT systems that can be applied across different types of devices. This algorithm is comprised of both unsupervised and supervised areas of machine learning combining the strongest facets of each method. These are the speed of unsupervised, as well as the accuracy of supervised machine learning. The algorithm involves the initial unsupervised k-means clustering of attacks. The k-means clustering algorithm groups the data as either DDOS, backdoor, ransomware, worm, trojan, password, and normal and assigns them to their clusters. Next, the clusters are then used by the AdaBoosted Naive Bayes supervised learning algorithm in order to teach itself which piece of data should be clustered to which specific type of attack. This increases the accuracy of the proposed algorithm by adding clustered data before the final classification step, ensuring a more accurate algorithm that can effectively classify attacks. The correct identification percentage scores for this proposed algorithm range anywhere from 90% to 100%, as well as rating the proposed algorithm's accuracy, precision, and recall on different datasets. These high scores achieve an accurate, flexible, scalable and optimised algorithm that could potentially be utilised in different IoT devices, ensuring strong data integrity and privacy.

## 1 Introduction

Detecting anomalies within Internet of Things (IoT) devices has become an increasingly important aspect of cybersecurity due to the growth in prevalence of these attacks (Wang, 2020). This growth has been fuelled by the current covid-19 pandemic, as large numbers of employees are working remotely from home. The increase in time spent at home means IoT devices are therefore used more, generating larger amounts of data, and subsequently becoming more attractive to malicious users. With the subsequent attractiveness of IoT devices increasing, a greater need for information security and specifically anomaly detection is required. Anomaly detection (AD) typically utilises supervised machine learning (SML) algorithms in order to detect intrusions.

## 1.1 Background

IoT refers to an ecosystem of interconnected devices that have computing and networking capabilities embedded within the object, in order to carry out several tasks (Sethi & Sarangi, 2018). The overall vision of the IoT is to add value, personalise the user's experiences and interactions with various "things". The IoT enables large-scale technological advancements in many different areas such as agriculture, smart cities, health and fitness, traffic, retail and logistics. IoT is often seen as a global infrastructure that enables the connectivity between the cyber and physical worlds based on existing structures, frameworks as well as previous, more basic IoT devices (Sethi & Sarangi, 2018). As IoT devices rely on a connection to each other and the Internet, this means that they are potential targets for malicious users, and therefore require measures to prevent and detect intrusions within a connected network.

AD within networking is utilised in order to ascertain behaviours deviating from the norm (Khamparia et al., 2020). An example of networking anomaly detection's actions would be the identification of suspicious networking packets from IP addresses flagged as malicious or the abnormal amount of traffic sent to and from internal or external networks. These attacks often take place within or against components of a network. These traditional networking components (firewall, router and switch) rely on AD algorithms, however networking for IoT is more difficult, as their security requirements differ due to the size, scalability and resources of connected devices (Rawat & Reddy, 2017). 6

The current IoT security measures are based on the principles of confidentiality, integrity, authorisation and availability. These measures include specific protocols for connecting and communicating with and through the internet, as well as machine learning algorithms for anomaly detection (Radoglou et al., 2019).

Examples of current security protocols for IoT are MQTT, CoAP, and AMQP. The MQTT (Message Queuing Telemetry Transport Protocol) model operates similarly to the HTTP, request-response operation as opposed to the MQTT mode that uses a publish-subscribe model. The publish-subscribe model operates by the publisher, first collecting the entity's data from different sensors. The subscriber subscribes to the sensors data which is displayed to the end-user. The third stakeholder within MQTT is the broker, which operates as the go-between, delivering the data from the publisher to the subscriber (Patel and Doshi, 2020). CoAP (Constrained Application Protocol) operates like the MQTT model however it relies on a URI (universal resource identifier). The CoAP model publishes data gathered to the URI, and the subscriber (user) subscribes to the resource indicated by the URI. When new data is published to the URI, the other users are notified. AMPQ (Advanced message queuing protocol) operates by utilising both the publish-subscribe and the request-response models. The AMPQ model communicates by requesting the user or publisher create and then broadcast that exchange. Subsequent communications take place by either the broad-caster or user utilising that name within the exchanges. At the same time, this is occurring, the user must create a queue and attach it to the exchange, with messages then matched to the queue in order to satisfy authentication. These security protocols operate concurrently with different AD methods in order to secure information and prevent malicious users from gaining access to IoT devices (Naik, 2017).

There are numerous current methods employed by IoT devices in order to ensure data and device integrity. These include but are not limited to authentication and access control, attack detection and mitigation, anomaly intrusion and detection and malware analysis (Hussain et al., 2020). These methods rely on different machine learning (ML) techniques to operate. Authentication and access control rely on artificial neural networks (ANN) and long short-term memory (LSTM). Attack detection and mitigation require a support vector machine (SVM), as well as deep learning autoencoders and K-nearest neighbors (KNN). Malware analysis can use recurrent neural networks (RNN), principal component analysis (PCA) and convolutional neural networks (CNN).

Intrusion and anomaly detections are sometimes composed of k-means clustering, decision trees (DT), and Naive Bayes (NB) (Hussain et al., 2020). Unsupervised machine learning (UML) algorithms, although more flexible regarding data absorption, require far less manual handling than SML (Usama et al., 2019).

Combining both UML and SML algorithms gives birth to hybrid machine learning (HML). HML involves utilising information associated with both ML strands and their strengths whilst minimising their weaknesses. An example of this is if there is an issue with a classification problem associated with SML, then using additional data points from UML could help with the labelling. Conversely, assisting a clustering method with increased knowledge relating to certain data points belonging to the same class would strengthen the use of UML clustering methods (van Engelen & Hoos, 2020).

AD methods operate concurrently on IoT smart devices, utilising ML algorithms. IoT devices generate different types of data, these include but are not limited to network and sensor data. This thesis will utilise telemetry/measurement data from sensors placed within several different IoT devices. This is done in order to further examine the usage of sensor data, versus the standard networking data that is based on TCP/IP or packet flow readings.

## 1.2 Research Questions

The main research topic this thesis seeks to answer is:

1) How much better would a hybrid machine learning algorithm comprised, of k-means clustering and Naive Bayes be than traditional SML algorithms when it comes to AD within IoT sensor data?

In effect, this thesis suggests that it is indeed a much better prospect than KNN, Random Forest or a regular Naive Bayes algorithm and measures it.

There are several sub-topics featured within this thesis as a result of the overarching theme. These relate to accuracy, scalability, speed and flexibility. These sub-questions are:

2) Does the proposed algorithm have higher accuracy, precision and recall scores than traditional SML methods?
3) Does the proposed algorithm have faster train and test times than the traditional SML algorithms?
4) Does the proposed algorithm maintain its strength as it operates on larger datasets?
5) Is the proposed algorithm able to be applied to different types of IoT devices?

The question of accuracy and speed also raises another question that could be answered within this thesis.

There is a suggestion that having an AD algorithm that is accurate but too slow is inefficient, compared to one that is slightly less accurate but quicker. This thesis seeks to ask and answer:

6) Does removing the highest-ranked subset of attributes to the type of anomaly, negatively impact the strength or speed of the algorithm?

Discussing the question regarding the highest-ranked subset of attributes allows for a closer look at the trade-off between speed and high accuracy. Another way of phrasing this question is, if you have an AD algorithm that is 90% accurate but takes 3 minutes to operate, versus an algorithm that is 87% accurate, that takes 30 seconds to predict, which algorithm is the better choice for the device? Answering the questions and sub-questions will allow this thesis to answer whether the HML k-means and Naive Bayes algorithm is indeed a viable alternative to traditional SML. This answer is based on the notion that having an AD algorithm that is slow, inaccurate, inflexible and unscalable would not be an effective use of resources within an IoT device.

## 1.3 Chapter Overview

The below chapters will be further broken down respectively: literature review, proposed algorithm, evaluation, results, discussion and analysis, conclusion and future work. The literature review will focus on discussing previous academic papers including providing a brief overview of IoT sensors, contrasting IoT networking data, anomalies, anomaly detection, both strands of machine learning, as well as then going on to discuss the usage of k-means and Naïve Bayes (NB) HML algorithms within academia. This thesis will then discuss how these features detect anomalies and the subsequent different types of anomalies. After the literature review, the new algorithm section details the requirements of the proposed algorithm, and how this proposed algorithm answers the research questions. The evaluation of the algorithm involves the discussion of the type of method used, as well as the subsequent hypothesis, the metrics for measurement as well as the process, and what is being measured. The datasets will also be detailed including their components, different sizes and types of IoT devices that were tested, different anomalies within each dataset and the metrics that will be used to measure the results will be evaluated. The Results section will address each research objective and question, as well as discuss the advantages of new algorithms and their trade-offs. The Discussion and Analysis section will review the research questions and give specific insights. Lastly, the conclusion will summarise the entirety of this thesis. The future work section discusses further study that could be explored at a later date.

## 2 Literature Review

This chapter will present what other authors have discussed regarding IoT, before going on to discuss sensor data, IoT anomalies and the subsequent different types, anomaly detection and machine learning (both supervised and unsupervised algorithms). Lastly, this chapter will then discuss hybrid ML before then giving an in-depth review of the many uses of HML algorithms, and then finishing with discussing the uses of k-means and or Naive Bayes within AD algorithms, as well as highlighting the raps of research.

### 2.1 IoT Overview

Kassab & Darabkh (2020) give an overview of IoT by first suggesting that there are issues within IoT, these issues are distribution, interoperability, scalability, resource scarcity and security. They then go on to discuss the issue of distribution. Distribution refers to the gathering of data from various sources and subsequently the processing of it. There are associated issues with this, as there are many different IoT vendors, and as such, the different products may not be compatible. As the interconnectedness of devices is integral to IoT as a whole, any such issue that affects this has severe consequences. Kassab & Darabkh (2020) then discuss the next issue of interoperability, which is that different vendors devices might not work together to achieve a certain goal, hence why the characteristic of architectural interoperability is necessary. This can be achieved by ensuring that each protocol and sensor permits other different vendors from reading and accessing their accumulated data. As there are billions of devices within the IoT environment, the amount of data that is generated is large, the applications within need to be designed with the ability to be scalable enough to process said data. Kassab & Darabkh (2020) discuss resource scarcity. Resource scarcity is the issue that relates to smart devices and the idea that they are resource-constrained, which means that they are limited by computation and energy requirements. Finally, they discuss security. Security is arguably the most integral architectural conceptual issue within the IoT sphere.

The lack of security for IoT devices will hurt their deployment as an institution. To address these issues Memon, Li, Nazeer, Khan, & Ahmed (2019) proposed other layers be added in order to address the drawbacks of the current IoT paradigm. These are the IoT Layer, the Fog Layer and the Distributed Cloud Layer. The Fog layer refers to the layer above the IoT layer that bridges the gap between smart objects, storage services and large-sized cloud computing servers. This is done by the fog layer utilising smart devices such as gateways, routers and dedicated computing devices by picking up some of the processing requirements, thus reducing network latency and workload from other smart objects. Memon, et al. (2019) then discuss the distributed cloud layer and the idea that this layer of the IoT sphere handles multiple processing of huge units, like a server rack. This is done once again with the aim of increasing the scalability of IoT devices by reducing memory and power consumption from smart devices. They suggest that this reduction is achieved by firstly an IoT, fog and cloud computing node sending and receiving information with one another. Next, the IoT node transmits the sensed data directly to the fog node belonging to the domain application. The fog node then either processes the data directly or passes it onto another node in the same domain. This occurs back and forth, allowing for the distribution of processing power requirements between different nodes.

## 2.2   IoT Sensors

Plageras et al. (2018) state that IoT Sensors consist of small devices which are wirelessly interconnected with one another, these are referred to as nodes (Plageras et al., 2018). Plageras et al. (2018) go on to suggest that they are called sensors, because they can collect information from the immediate environment such as a smart home that has a smart thermometer, or an automatic light that uses sensors to see if there is someone in front of it. IoT sensors are also seen to take continuous measurements at different points throughout operation allowing for constant monitoring.

## 2.3   Contrasting IoT and Standard Networking Data

There are several differences between IoT and standard networking data. Alsaedi et al. (2020) suggest that the current networking anomaly detection datasets, that are based on current systems primarily contain packet-level and flow-level information. This data is handy in detecting attacks on the network but fails to address the issue of attacks that specifically aim to change sensor data or manipulate IoT devices. Standard networking data found in basic datasets such as NSL-KDD does not contain information gathered from the actual sensors. Data gathered from the sensors, is relatively simple in contrast to the packet-level and flow level information previously mentioned (Alsaedi et al., 2020). Alsaedi et al (2020) suggest that this is a major gap in academia right now and that they attempting to address this.

## 2.4   Anomalies

Quek et al. (2020) discuss anomalies by first giving a base definition. According to the authors, anomalies are simply a deviation of normal conditions from the operating paradigm. They then go on to state that the two main assumptions regarding anomalies are that they occur rarely and that they are distinguishable from normal data. Looking specifically at the concept of anomalies within the IoT space, this typically refers to malicious incursions into the network or the IoT device (Sahu & Mukherjee, 2020). Sahu and Mukherjee (2020) then continue, stating that as there many different types of IoT devices within the ecosystem, therefore logically the types of anomalies also vary. This then brings the authors to discussing zero-day attacks, stating that zero-day attacks (ZDA) are numerous, and in effect are unknown anomalies that can be found within IoT devices. A ZDA attack refers to an event of involving different types of anomalies, including but not limited to denial of service, malicious control, malicious operation, scan and spying attacks. Sahu et al then go on to list the types of anomalies, as well as background information and examples. They suggest that denial of service (DoS) attacks are like the attacks found within traditional networking. DoS attacks operate by making the required service crash or become unresponsive. This is done by ensuring that there are no allocated resources left to respond to the constant requests. The data type probing anomaly occurs when the device receives data that has been changed, an example of this is if a sensor is expecting an integer value, but instead receives a string. Malicious control refers to another unintended user attempting to gain control of the network traffic. Malicious operation occurs when the original activity is obfuscated and hidden. Scanning anomalies refer to when a malicious user intends to replicate the client or server credentials in order to gain access to user data. The spying anomaly involves when eavesdropping occurs, listening in on specific areas of the network to discover important information (Sahu & Mukherjee, 2020).

## 2.5 Anomalies within this Thesis

The types of anomalies that are found within this thesis are Scanning, distributed denial of service (DDoS), ransomware, backdoor, injection attack, cross-site scripting (XSS) and password cracking attacks. Alsaedi et al. (2020) describe the types of attacks in detail. The authors start by stating that scanning attacks refer to the first step a malicious user takes, in order to attempt to gain access to the network system. Scanning attacks involve information gathering, targeting areas of vulnerability such as open ports and available services. DDoS attacks involve the malicious user flooding the victim with requests to disrupt access to services. They are often launched by many compromised devices known as botnets or bots. These compromised devices flood the target, often overwhelming their memory and bandwidth. This is particularly dangerous for IoT devices as they have limited computational power and storage capacity. Ransomware attacks are often malware-based and usually operate by denying the user access to a system or specific services. The malicious hackers will then sell decryption software back to the victims in exchange for returned access to their system. Backdoor is a passive type of attack, in which the bad actor will attempt to gain access to the victim's system remotely, usually with malware. These compromised systems often perform part of a botnet to launch DDoS attacks. Injection attacks involve executing snippets of malicious code or data into targeted applications. The injection attack can manipulate telemetry data and control commands in order to disrupt normal operations. XSS attempts to operate malicious code on a web server that is connected to the targeted device. The XSS allows the malicious user to inject scripts of coding which can then compromise authentication procedures between different devices, and the webserver. The Password cracking attack occurs when an attacker uses cracking methods to guess a password and subsequently gain access to the attacked system. This type of attack can allow attackers to bypass authentication methods (Alsaedi et al, 2020).

## 2.6 Anomaly Detection

Tsai et al. (2009) state that anomaly detection within IoT comprises different areas, with differing methodologies. The authors then discuss the differing anomaly detection methodologies. The first anomaly detection method is classifying either signature-based or semantic-based. Signature-based AD refers to detecting attacks through threats or signatures that are already known. Anomaly-based detection schemes operate by employing statistical, machine or protocol-specific information and then building a model featuring legitimate traffic. This model is then used as a reference point to classify either normal or abnormal traffic. There are also hybrid systems that combine both detection and classification methods according to Lawal et al. (2020). Other anomaly detection methods can occur in real or non-real-time, real-time referring to occurring synchronously, and non-real-time meaning asynchronous. Anomalies can be found within either the actual network flows of IoT devices or they can be discovered by examining sensor information (Tsai et al., 2009). Some of the common types of machine learning and anomaly detection algorithms are discussed below.

## 2.7 Machine Learning

Bengio et al. (2015) discusses AD machine learning methods and then breaks them down into SML and UML. For SML, the training set functions as samples of input data points, in conjunction with corresponding appropriate target vectors (labels). This contrasts with unsupervised learning, which does not require the use of labels. The main objective of SML is to learn how to predict the output data from a given input vector. Classification is a method widely used and consists of the target labels known as classification tasks examining a finite number of discrete categories (Bengio et al., 2015). This thesis will detail SML algorithms, and then discuss UML in the next part of this literature review.

## 2.8 Supervised Learning

The SML algorithms that will be discussed below are K-Nearest Neighbours (KNN), Naive Bayes (NB) and Random Forest (RF). These are chosen as they will be the traditional SML algorithms that our proposed algorithm will be tested against.

**K-Nearest Neighbors.** Jagdish et al. (2005) go into detail regarding KNN. They suggest that the main goal of KNN is to classify a new, discrete data point by finding the K-given data points in the training set. This is done by examining the data points closest to the input or feature space. To find the KNN, a measure of distance metric (Euclidean distance, L∞ norm, angle, Mahalanobis or hamming distance) must be utilised. Jagdish et al. (2005) state that in order to solve the problem, first let the new data point (input vector) be seen as x, its K Nearest Neighbors by Nk(X), the predicted class label for x by y, and the specific class variable by t (a discrete random variable). Furthermore, 1(.) denotes the indicator function: 1(s) = 1 if s is true and 1(s) = 0 if not. The input data point (x) will be generated by the mode of its neighbour's labels (Jagdish et al., 2005).

The authors then go on to depict KNN in the following way:

KNN is mathematically depicted as:

$$p(t = c | x, K) = {1}/{Ek} \sum_{i \in Nk(x)} 1(ti = c)$$

$$y = \arg\max p(t = c | x, K)$$

Like every other type of ML algorithm, there are both pros and cons. A downside of KNN is that it requires the storing of the entire dataset, meaning the more data there is, the more that must be stored by the algorithm, meaning it is not a strong choice for larger datasets and is not as scalable as others. This means that KNN is not the most scalable algorithm and would therefore not be optimal for use in IoT devices.

**Naïve Bayes.** Zhang (2005) discusses NB. They start by stating that NB's primary function is to apply the Bayes theorem with the naive assumption of independence between the features (attributes) of z when given the class variable t.

Zhang (2005) starts by stating denoting the input vector z = ($Z_1,.....Z_M$).

When applying the Bayes theorem, the below formula is used:

$$p(t = c) | Z_1,.....Z_M) = \frac{p(Z_1,.....Z_M | t = c) p_{(t=c)}}{p(t = c | Z_1,.....Z_M)}$$

Then subsequently integrate the naive independence as well as the subsequent simplifications, which leaves us with:

$$p(t = c | Z_1,.....Z_M \propto p(t = c) \Pi_{j=1}^{M} \ p(Z_j | t = c)$$

The way in which the classification takes place is shown below.

$$y = \arg m_c ax p(t = c) \Pi_{j=1}^{M} \ p(Z_j | t = c)$$

It should be noted that y shows the predicted class label for z. There are also different types of classifiers using different methods of distribution to estimate *p(t=c)*. NB already has a strong foundation within the AD sector, its uses already include spam filtering and text classifications (Zhang, 2005)

**Random Forest.** Wang et al. (2018) discuss the basic methodology of RF. They start by discussing the positives of RF and suggesting that it has a strong ability to handle various types of data, and as such has a wide area of application within academia. Wang et al. (2018) then describe the algorithm by first stating the assumptions. These assumptions are a data set is annotated as Dn with n being the instances (X, Y), it should also be noted that, $X \in \mathbb{R}D$. This approach combines numerous decision trees, independently trained to form a forest. It should also be noted that each tree is a partition of the data space. This means that as $\mathbb{R}D$ is the full set of data, then a leaf is a partitioned subsection of the entire dataset, and each node corresponds to a cell of data space. The methodology of the RF is presented below.

1. At the start of the tree construction, n sample points are taken from the Dn dataset. Only these samples are used to construct the tree.
2. Tree node mtry features (mtry < D) are then randomly sampled from the original D datasets. These samples are then used for the selection of the splitting point. After one split has occurred, the algorithm continues to split repeatedly until the stopping condition is met.
3. RF's then average the result from each tree (Wang et al., 2018).

## 2.9 Unsupervised Learning

Usama et al. (2019) discuss UML generally before going onto data clustering. They start by stating that UML allows for the analysis of raw data, thereby helping in generating insights into unlabeled data. UML has many different applications within the ML sphere. UML algorithms are utilised in areas of speech recognition, computer vision. Furthermore, they are viewed as flexible and scalable in nature, it could therefore be suggested that these uses could be then transferred to AD. Usama et al. (2019) also suggest that due to these strengths, they could then be applied to areas within network management, monitoring, and optimisation. UML techniques can be divided into different sections, these include but are not limited to hierarchical learning, data clustering, latent variable models, dimensionality reduction techniques and outlier detection (Usama et al., 2019). This thesis will specifically discuss data clustering below.

## 2.10 Data Clustering

Usama et al. (2019) state that data clustering encompasses the organisation of data in natural, meaningful groups (clusters) based on the high similarity between different features. Clustering, therefore, attempts to find hidden patterns within the input, unlabeled vectors. The clusters are also organised in such a way that promotes high intra-cluster and low inter-cluster similarity (Usama et al., 2019). The authors then suggest that clustering is widely applied to many different disciplines, these include but are not limited to ML, data mining, network analysis, pattern recognition, and anomaly detection. Data clustering can be further broken down into 3 areas, Hierarchical Clus-tering, Bayesian Clustering, and Partitional Clustering. We will look specifically at partitional clustering as it precedes K-means (KM) clustering, the algorithm used examined within this thesis.

**2.11 Partional Clustering**

Usama et al. (2019) discuss partitional clustering generally and then go on to discuss the pros of this method. They start by stating that clustering is a method of organising data into a set of disjointed clusters. Partitional clustering has an advantage over other types of anomaly detection algorithms in that they can incorporate knowledge relating to the size of clusters, by relying on the specified distance functions which ensure accurate shape generation. Frigui (2005) mentions that there are several drawbacks to partitional clustering. These are the difficulty in determining the number of clusters, the predisposition to being negatively impacted by outliers and data noise, and the issue of cluster initialisation. Partitional clustering can be further broken down into K-medoids, expectation means, and k-means. As K-means was the chosen partitional clustering type, this is the algorithm that will be discussed below.

**2.12 K Means Clustering**

Zhao, Deng, and Ngo, 2018 start by giving a general overview of the KM clustering algorithm. KM clustering is defined as an iterative expectation maximisation approach. It operates by including 3 steps. The first step is to initialise the k cluster centroids, the next is to assign each sample collected to its closest centroid, and the last step is to reorganise the cluster centroids with the assignments computed in step 2, and then repeating step 2 until convergence is met. These steps are repeated until the centroids of the clusters do not change between consecutive iterative rounds. Zhao et al then go on to display the clustering procedure. This is done by stating that $\{x_l \in R^d\}$ $i=1\ldots n$ are samples that are required to be clustered and $C\ldots k \in Rd$ is the cluster centroids. The above function represents the iterative and clustering steps. Before the specific KM algorithm is depicted, there are several rules that need to be specified.

Qi, Yu, Wang, Liu & Wang, 2017 describe the KM algorithm. This was done by first stating several assumptions. One assumption is that the given dataset is denoted as D, and then $D=\{pi|i = 1,\ldots,n\}$, pi found in d-dimensional space. The first step (seeding) begins by selecting k clusters, by minimising the sum of squared errors (SSE). The first equation depicted shows the operation of the KM clustering algorithm. This algorithm is found below:

$$\boldsymbol{SSE} = \sum_{j=i}^{k} \sum_{i=1}^{n} \delta_{ij} \| p_i - m_j \|^2$$

$(\delta_{ij} = 1 \; if \; p_i \in C_j \; and \; 0 \; otherwise)$

It should be noted that where $\|pi-mj\|$ is shown, this indicates the distance between the point $pi$ and the cluster $Cj$, as well as its cluster centre $mj$. This is shown below in the subsequent equation.

$$m_j = \frac{\Sigma_{pi}\in C_j P_i}{|C_j|}$$

(Qi et al., 2017). The combination of the previous supervised and unsupervised machine learning algorithms leads to the discussion regarding hybrid machine learning HML.

### 2.13 Hybrid Machine Learning

Li et al. (2018) discuss HML algorithms. The authors suggest that these algorithms consist of two forms of machine learning, unsupervised and supervised. Both unsupervised and supervised machine learning algorithms have strengths and weaknesses. It is suggested that utilising a combination of these two types is an effective way of combating each other's specific weaknesses. There are numerous examples of hybrid machine learning algorithms. A popular way is to utilise one UML as the data aggregator, and then another one as the classifier. This is what has been proposed, a clustering algorithm to gather data, and then a supervised classifier to classify the data.

After the basic overview of IoT and the different types of algorithms has been discussed, this thesis will now detail the potential uses of hybrid ML in the papers below.

### 2.14 Uses for Hybrid Machine Learning Algorithms

One potential use for HML refers to the integration of KM and NB for smart air conditioning (AC) monitoring and control in WSAN networks, as proposed by Kristianto et al. (2019). Kristianto et al. (2019) describe the combination of KM and NB, and the way it operates. This is done by utilising the classification of the NB algorithm to determine the operation of the AC. The sensors generate the unsupervised dataset, which is then clustered and formed into a supervised dataset. This supervised dataset is then passed on to the NB classifier and the subsequent instructions are then passed onto the controlling server. New data is then assigned to the specific clusters based on the previously NB classifications of previously collected data. After that data is received by the server, it is then passed onto the remote actuator, which then determines the operation of the AC. Kristianto et al. (2019) then show that their combination, in terms of results, is rated by accuracy, precision, recall and error rate as 90%, 83%, 100% and 10% respectively. However, the conclusion presented is basic and suggests that merely because they have strong scores, their algorithm is good, there is a lack of reasoning behind this claim.

Wayahdi et al. (2020) suggest a combination of KM and a NB classifier to classify an image. This is done by attributing numbers to sections of an image based on the characteristics and statistics of the picture, as well as the hue, saturation, red, green, blue, kurtosis and skewness. Wayahdi et al. (2020) provided an example and stated that this was first done with an image of a banana which was then resized to 100x100 pixels. Then the image is extracted into different numeric characteristics, based on the attributes. The grouping occurs using KM clustering. The classification via NB is repeated for different images each with different centroids. The results for this state that a total correct percentage of 85% was found. It could be suggested that the 85% accuracy rate could be fractionally low, particularly when using an SML algorithm. (Wayahdi et al., 2020).

Ali et al. (2016) suggests a way in which to utilise KM and NB as well as feature selection in text document categorisation. The hybrid machine learning algorithm would be utilised by first pre-processing the documents before clustering by removing redundant and duplicate words, question marks and conjunctions. Next, the feature selection phase refers to the operation of the proposed model that involves further pruning. After this, the KM algorithm is called, which then calculates centroids of clusters, the clusters themselves as well as the minimum distance function. The minimum distance function denotes which features are allocated to the nearest K cluster. Next, the optimisation by the NB algorithm creates a specific cluster according to the predicted probabilities. The optimisation process continues as new centroids are created with new documents; this keeps occurring till all the allocated documents are clustered. The results of this algorithm suggest that the combined KM based NB is an accurate algorithm over the 4 chosen datasets. This is evident as for the first dataset, the proposed algorithm received 91.60% purity and 72.20% entropy for the proposed model compared to the 86.80% purity score and the 77.00% entropy score. Entropy refers to the measure of quality for clustering. However, the authors chose how many k-clusters simply by trial and error, they did not demonstrate any method. This could have been improved on by mathematically suggesting the strongest possibility or using an expectation means algorithm and automatically assigning the number of clusters as per the features within the dataset.

Fadhil (2021) proposed an algorithm that operates on the hybrid KM and NB systems to predict the performance of an employee. In the first phase, the KM algorithm begins the clustering process in order to determine the training data. This data includes classes (excellent, very good, good, average, and bad). The Euclidean distance, using the class data measures the distance between data and the first centroid of the algorithm. After the initial centroid initialisation and Euclidean distance calculations occur, the class for ever cluster is then analysed, and the average data for each variable within said cluster is found. This keeps occurring till all the data has been clustered. The average data score for each variable is compared to the centroid, and if it is not equal to the centroid's value, then the distance calculation is repeated until the average data is equal to the value of the centroid for each variable. After the value has been chosen, the NB classifier is used. With the best results of each centroid value being output to the user. A critique that can be seen is that this algorithm relies heavily on there being no outliers within the data, as the centroid value of this algorithm plays a large part, as the process repeats itself till the centroid value is the same. As KM is outlier sensitive, it could be seen that any outlier or null data would affect the centroid, making the algorithm keep repeating the initialisation stage (Fadhil, 2021). This appears to be a systemic floor within the methodology and could be addressed by firstly pre-processing and removing the obvious outliers or null values.

This thesis will now discuss using KM and NB for AD within the field of networking. These efforts will be discussed, reviewed and critiqued below.

## 2.15 Hybrid AD algorithms using KM and Naïve Bayes

There are several HML anomaly detection systems that have been proposed. One such AD system is that of a hybrid KM clustering and NB classification technique. This paper suggests potentially using a J48 technique combined with a correlation-based feature selection (CFS) methodology as well as a KM and NB algorithm. For this literature review, only the KM and NB will be assessed. This AD system was suggested by authors Bagui et al. (2019).

Bagui et al. (2019) provide an overview of the way the experiment is designed and then discuss the results and any issues that have arisen. Firstly, the authors state that the algorithm operates by ingesting 8000 random records from the UNSW-NB15 dataset. Feature selection is then performed using the KM clustering and CFS. CFS evaluates the benefits of each subset. This was run on each attack family (fuzzers, analysis, backdoor, dos, exploits, generics, reconnaissance, shellcode and worms). Once the features were selected for each of the attack families, the classification step began. The classification step involves the usage of 2 different algorithms, the NB and the J48 decision tree. The results showed that NB produced the best rates of classification, coming in with 80.03, 90.66, 90.02, 92.97, 46.70, 92.61, 71.42, 75.24 and 99 for fuzzers, analysis, backdoor, dos, exploits, generics, reconnaissance, shellcode and worms respectively with CFS. Contrasted to without CFS, the scores were 57, 74, 66, 66, 56.69, 83, 65, 72, and 84 respectively. Bagui et al. (2019) found that there were substantial increases in accuracy for the detection rate related to the use of CFS. Our proposed algorithm differs from the authors algorithm in several ways. The main difference is that we are not using feature selection as the core of the method. Instead, we are using CFS to rank the feature with the highest correlation ranking to the label, and then using that ranking to remove the highest ranked feature. This is done to see if the training and testing time can be improved whilst maintaining accuracy. This is due in large to the data we are analysing, and the subsequent implementation methods. Since our data is not as complex, adding in feature selection would be redundant. Our algorithm is designed to be used on simpler data gathered from IoT sensors. This allows for a more agile and less processing intensive approach, with comparable if not more accurate scores.

Bhatt & Thakker (2019) suggested an algorithm to aid in the removal of botnet attacks using an ensemble classifier within IoT devices. Bhatt & Thakker (2019) suggested collecting hacker activity patterns from the IoT devices as opposed to traditional usage of network statistics. Next, the modelling of the attacking information takes place within a tree-based structure by stacking the classifier. Lastly, the attacks are then clustered by a protein similarity algorithm (PROSIMA). The similarities to our proposed algorithm are minor, in so far as this utilises a clustering algorithm, AdAboosting, as well as a classifier and IoT device data. It should be noted that that algorithm does in fact involve the use of feature selection and is more complex in nature. As mentioned above, our proposed algorithm appears to be less complicated in nature, which fits with the design ethos of speed due to the resource-constrained nature of IoT devices (Bhatt & Thakker, 2019).

Om & Kundu (2012) proposed a hybrid intrusion detection system that combines KM and two additional classification algorithms, KNN and NB. It consists of feature selection based on entropy evaluation operating on the KD-99 dataset (Om & Kundu, 2012). Om & Kundu (2012) go on to discuss the method of operation for their proposed NB intrusion detection system (IDS). The IDS starts with first applying KM clustering to the dataset, specifying the number of clusters into either normal or anomalous clusters. The number of clusters is set to 5 (user 2 root, remote to local, probe DoS and normal). The data is then separated into two parts, one part for testing and the other for training. In the training phase, the labelled records are assigned to the K-Nearest Neighbors. The KNN is then trained on these. The subsequent rest of the data is then passed through the KNN classifier, it should also be noted that this method involves feature selection. This algorithm operates similarly to the proposed algorithm, with the clustering into classification. This delivered strong results however, there was no mention of the training and testing time. Moreover, as KNN needs to ingest the entire dataset, this further reinforces the theory that this algorithm was particularly slow in comparison to others.

Sharma et al. (2012) suggest an improved intrusion detection technique based on KM clustering via the usage of NB as a classifier. Sharmai et al. (2012) state that the algorithm consists of several steps. Firstly, it begins by pre-processing (feature selection) and normalisation. Next, the KM clustering algorithm is run, followed by the classification via NB. After this occurs, the testing and validation of the performance take place, with the results displayed. Once again, this model relies on the use of feature selection before the algorithm occurs, in order to assist with getting higher accuracy rankings. Furthermore, they do not utilise any boosting methods. This could potentially increase their accuracy, therefore reducing bias and variance and allowing for a stronger overall algorithm (Sharma et al, 2012). It can also be noted that Sharma et al suggest that the design of their algorithm is meant to be a general algorithm, however, the results suggest a different picture. The results demonstrate that in fact it is accurate, it generates more false positives, as well as being less accurate for 2 out of 5 types of attack, and on par with the other baseline AD algorithm. This means that the aim of this algorithm is largely unmet, as only 2 attack types were detected consistently.

Soe et al. (2020) suggest a lightweight, sequential attack detection architecture that is based on 4 modules specifically to be used on IoT devices. The 4 modules are data collection, data categorisation, feature selection and model training. The first module data collection ingests benign and attack data from the network and IoT environment. Module 2 categorises each piece of attack and benign network data. Furthermore, each category includes all data with the same attack class and benign data. The feature selection module designates the highest correlated features of each class, as there are 8 types of attack class, the feature selection module will select features for these classes. The model selector module places and then evaluates several different ML algorithms, selecting the more accurate one. The attack detection occurs afterwards and involves feature extraction and then alert generation. A critique offered for this paper is that this algorithm claims to be lightweight, however, it is itself quite complex. This is evident as there are 4 modules required for this algorithm to operate, one of them being feature selection and as stated above is memory intensive. Moreover, the claim of lightweight is also contradicted by the operation of said algorithm. As the algorithm requires data collection, data categorisation, and feature selection to occur one after the other. After the feature selection occurs, the pre-processing of the attack detector begins, which then further involves attack detection and the subsequent alert generation, all this adds to the inherent complexity of this algorithm (Soe et al., 2020).

Samrin and Vasumathi (2018) suggest an algorithm that is a combination of the KM clustering algorithm and an artificial neural network (ANN). This algorithm is broken down into the training phase and the testing phase. The training phase involves firstly completing the KM clustering method above. Once the clusters have been found, each output cluster from the KM clustering operation is trained by the ANN. This step keeps repeating for each neural network node till the output is produced, the training phase occurs afterwards. The training phase involves the total usage of the data, redistributed back through the KM algorithm, as well as the ANN in order to learn through itself. After the self-learning has taken place, the algorithm's accuracy, sensitivity, and specificity are returned. The accuracy scores were dependent on the number of clusters that were found within the data. Cluster sizes of 10,15,20,25 and 30 had accuracy ratings of 88,89,92,88 and 89 respectively. The sensitivity (probability the algorithms predict positive examples) was found to be 80,83,76,73 and 83, once again depending on the cluster size. Lastly, the specificity rating (probability algorithms can predict negative examples) is rated as 66,68,68,69 and 59. It should be noted that the accuracy was better in every cluster, as well as the sensitivity. The specificity measurement proved to be negligible, with neither good nor bad results (Samrin & Vasumathi, 2018).

Saputra et al. (2018) used a combination of NB and KM to aid with the classification of illiteracy. This is done in several steps, research data, pre-processing, clustering and classification and lastly the testing method. The first step operates by gathering the needed data from different sources, such as high and elementary schools, as well as ascertaining characteristic data such as unemployment rates and education enrolment percentage and illiteracy rate. After the research data step, pre-processing takes place and involves the combining of the accumulated data into 1, assessable table. The data is therefore pruned to only include related data, and then standardised so it is usable for mining purposes. The third step, clustering and classification, occurs next, with the KM being implemented, first forming 2, then 3, then 5 clusters in order to ascertain the different illiteracy levels. This step is carried out in order to find the number of clusters that are considered optimal and can then be used in the classification step with the NB algorithm. In the classification process, the NB will be repeated 3 times using the training data, this will allow for the assessment of what types of illiteracy levels can be found. Lastly, the testing method involves utilising the k-fold method, in which 10 folds are used to validate results on the experimental data. After this is carried out, the results are analysed using accuracy and error rate as the metrics of analysis. The results for this indicate that the NB algorithm is a good candidate for the use of classifying clustered objects and finding anomalies. Furthermore, 3 clusters are ideal to be used. It should be noted that Saputra et al. (2018) did not use any feature selection algorithms, cutting down on operational time. They also got final accu-racy ratings of 93% upwards, for each run of the algorithm. This suggests that using a combination of KM and NB gives strong results, this once again reinforces the proof of concept that KM and NB together enable a strong detection algorithm (Saputra et al., 2018).

Varuna and Natesan (2015) continue this same path with focusing on combining NB and KM together for an anomaly detection algorithm, this time specifically on the networking data involving the NSL-KDD set. This paper's algorithm is broken down into 3 stages, clustering, calculating distance sum, and classification. The first step involves clustering utilising the KM algorithm and grouping the objects into similar groups. Next, the original dataset is transformed into a newer dataset including the previously clustered samples. The newer dataset includes the combination of the training and testing data, which was used to calculate the K distance sums for each sample. The classification step involves the training dataset, being used to construct an NB classifier. There are several issues with this paper and its subsequent results. Varuna et al discuss the metrics for evaluating the proposed algorithm, listing detection rate, false-positive rate, and accuracy. They then go on to not mention what these are for each step or give any other details regarding their results. The results that are provided indicate that this proposed algorithm scored lower in 2 out of 5 sections (the normal and dos predictions). Another issue that can be seen is that the way with this architectural framework is set out. The algorithm appears to be complicated in terms of operation, and the results not as strong. If something is used in AD, you would naturally expect stronger results the more steps that are involved, as this would indicate a more advanced thought process. This is not the case as mentioned above, the results would be considered average at best despite the intricate nature of the algorithm and as such suggest poor methodology and theory.

Tayal et al. (2016) propose a way to detect spam in mail servers utilising a modified KM algorithm as well as NB for classification, this is outlined in the 7-step process. The first step involves the establishment of the mail server. Next, the dataset is collected and then placed into two sections, the training data and testing data with a 60% and 40% breakdown respectively. Step 3 involves the pre-processing of data by partitioning two parts, the header and body of the email. The header has general information such as sender id, date, time, subject and internet service provider. The body of the email contains the message of the sender. The pre-processing step is completed to assist the algorithm with interpreting the database and as such assisting the readability of the data. The pre-processing step consists of feature extraction, dimensionality reduction, evacuation of stop words and stemming. This allows for tokenisation to occur. Tokenisation refers to the idea of representing the broken-down words as tokens, allowing for the algorithm to read and subsequently operate computations on them. Step 4 involves term selection, and revolves around the frequency of the associated token, in every database, the frequency of said token demonstrates how often a word appears. Step 5 involves the operation of a modified KM algorithm that segregates the body, the email body, into groups of similar messages based on the premise of comparability via the Euclidean distance. The modified KM eliminates the empty clusters expanding spam recognition. Step 6 relies on the NB classification and suggests based on probability whether the message is spam. Lastly, step 7 is the results step in which after computation occurs, the precision will be ascertained and displayed. Tayal et al. (2016) state that this proposed algorithm returns rates of 96% precision in the detection of spam which is good. However, there are no specific mentions of what a base NB, KM or modified KM gives. It should be noted that there is a graph that shows accuracy rates and that the NB is depicted as having ~78%, the KM as ~90% and modified KM as ~91%. It is hard to see how the authors have rated each algorithm specifically, it is not stated within the text (Tayal et al., 2016). The research gaps found within the texts identified that although, the KMANB has been used before, it has not been used on IoT sensor data. One way that KMANB has been used was for AD on networking data. Although this is the same purpose (AD) our method involves not using feature reduction as the core of the algorithm. This goes against traditional thinking AD thinking as feature reduction is a consistent theme found throughout the above papers. Our algorithm fills this gap, we use KMANB on IoT data, across multiple devices and with no feature reduction.

## 3 Proposed Algorithm

The Proposed algorithm has several requirements that need to be met, for it to be deemed a stronger alternative to the traditional AD algorithms. It should be noted that the below listed requirements are all needed to answer how much better would a hybrid machine learning algorithm comprised, of KM clustering and Naive Bayes be than traditional SML algorithms when it comes to AD within IoT sensor data?

1) Accuracy
2) Speed
3) Scalability
4) Flexibility

Each of these can also be used to answer the other listed questions. Accuracy is required to answer the question of does the proposed algorithm have higher accuracy, precision and recall scores than traditional SML methods? Speed is required to answer does the proposed algorithm have faster train and test times than the traditional SML algorithms? Scalability is required to answer does the proposed algorithm maintain its strength as it operates on larger datasets. Flexibility is required to answer is the proposed algorithm able to be applied to different types of IoT devices? The requirement of accuracy refers to the idea that the AD algorithm must be able to consistently predict the correct data. Speed refers to the training and testing time and whether it is quicker than the other traditional algorithms. Scalability means that it must be able to be used on different sizes of datasets, moving from a smaller dataset to a larger one. Flexibility means that it must be able to be used across different types of devices to a high standard.

### 3.1 KMANB

The KM Adaboosted Naive Bayes (KMANB) combines the strengths of the KM clustering algorithm, with the strength of the Naive Bayes SML algorithm. As men-tioned above, the KM clustering algorithm is particularly strong at data aggregation, meaning it can produce clusters quickly. This works in conjunction with NB as its training times are quick, and testing scores are accurate. The clustering works with the sensor data, as this adds additional, soft clusters to the dataset, ensuring that there is more data to be classified by the NB. If there are correctly clustered samples within the set IE. A normal packet is correctly clustered with the other normal data, then NB would theoretically learn off these and be able to adequately predict the other anomalies within the IoT sensor data. The KMANB algorithm is depicted below.

Fig 1. KMANB Algorithm

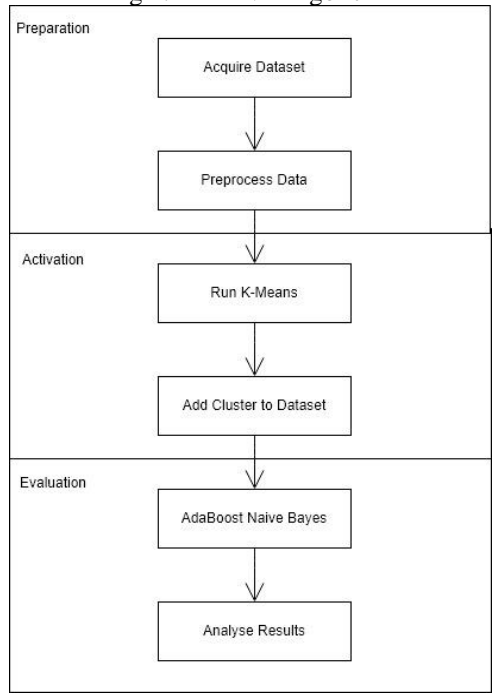

The way in which the algorithm operates is depicted above in Fig 1. It shows that the KMANB algorithm can be broken down into 3 steps. Step 1 is the preparation phase, step 2 is the activation phase and step 3 are the evaluation phase.

### 3.2 Step 1 – Preparation

Step 1 involves the selection of the dataset, as well as any data manipulation and pre-processing. This is seen as first, when the dataset was chosen, the labelling on it was changed, from a 1 or 0 to normal or anomaly. This was done in order to make it data pre-processing easier. Next, the pre-process step dictates that the data was normalised to ensure that the weights of different scales did not skew the data. It should also be noted that in some of the other datasets, another step was added to the pre-processing section to ensure that the clustering could be actioned. An example of this is with changing the string (Boolean) data of sphone_signal to nominal, in the IoT garage door train and test dataset. The use of Boolean was not allowing the clustering to occur.

### 3.3  Step 2 – Activation

The activation step refers to the initial operation of the KM clustering algorithm, the addition of that cluster to the set, and then the use of an Adaboosted Naive Bayes for the classification step. The activation of the KM algorithm includes specifying the number of clusters to be used, as well as any attributes to be ignored and the type of clustering to be performed and compared to a class of features. Firstly, the number of clusters that are specified relies on the different types of anomalies plus the one extra for normal. This can be represented as C (clusters) = A (anomalies) + 1 (normal). An example is that the IoT Fridge has 6 anomalies and 1 normal, therefore this would be denoted as C=7 or C = 6 + 1. Once the clusters have been set, the algorithm is run with a class to cluster evaluation, ignoring both type and label within the set. This ensures that the unsupervised element of this algorithm is maintained, and that the KM clustering looks at the data uninfluenced and then clusters accordingly. Lastly, the classification step happens when the Adaboosted Naive Bayes (ANB) is applied to the clustered data. The ANB uses the learnt clusters and then bases its predictions off that, meeting the supervised portion of the algorithm.

### 3.4  Step 3 – Evaluation

The third and final step is the evaluation portion of the algorithm. This refers to the algorithm being evaluated by looking at each of the clusters and their respective scores, then presenting an overall score. These will be discussed in a later section.

### 3.5  Algorithm Design

The design of this algorithm is largely based on the idea of maximising strengths and minimising the weaknesses of each type of algorithm. Furthermore, it was decided against any feature selection techniques within the main body of the algorithm, due in large to the dataset already being not big enough. This goes against a lot of traditional thinking regarding anomaly detection, where feature selection is regarded as an integral step in dimensionality reduction as well as increasing computational efficiency (Teh et al., 2021). In fact, we have done the opposite and have added more data to the set, this once again is in step with the trade-off with speed, precision and data size. It will be suggested that this addition has allowed for an increase in accuracy at the negligible cost of further clusters within the dataset.

## 4  Evaluation

The methodology presented below will demonstrate the actions being taken to address the need for an accurate HML algorithm, to be used within IoT sensor data. This will include examining the different datasets. It should be noted that the datasets were provided from Alsaedi, et al (2020).

### 4.1 Methodology

Our methodology involved firstly analysing and discussing many various texts within the academic sphere, to find a research gap. Once this gap was identified, 5 different datasets were researched in order to assess which, one would best suit an experiment regarding IoT smart devices and sensor data. The ToN_IoT Dataset was subsequently chosen. The ToN_IoT dataset is based on the new generation of IoT or IIoT (industrial internet of things) devices (IoT 4.0). ToN_IoT is used to evaluate different cyber security applications, as well as other artificial intelligence and, machine or deep learning algorithms. The ToN_IoT can be broken down into 4 different datasets; these are raw datasets, processed datasets, train and test datasets and security event ground truth datasets. The first KMANB algorithms were run on the base Train and Test datasets, to compare to the traditional SML algorithms and to examine whether the proposed algorithm was firstly accurate, and then to see if it maintained accuracy over the over devices. Once this was confirmed, correlation-based feature subset selection was run to discover the highest ranked feature to the anomaly label, and then remove it. Once removed, the experiments were run again to test whether the KMANB would maintain its accuracy without the strongest feature correlation related to anomaly type. After this, the processed datasets were manipulated and cut down slightly to make them operatable. These experiments were then actioned to see if the KMANB was scalable to larger datasets. The processed datasets were also run without the highest ranked feature relating to anomaly type, in order to further test investigate the potential for the trade-off between speed and accuracy (Alsaedi et al., 2020).

### 4.2 Train and Test Dataset

The Train_Test dataset consists of 7 different IoT devices which are a fridge, garage door, GPS tracker, modbus, motion light, thermostat and weather sensor. Each of these datasets had different sizes and different data types. Table 2 provides an overview of each dataset size, including the number of rows, columns and size of the file. tables 3,5,7,9,11, 13 and 15 give a profile of each IoT device. These device profiles contain the date, time, both labels (attack or normal and type of anomaly) as well as device-specific sensor information. Tables 4,6,8,10,12 and 14 depict the number of anomalies to be found within each IoT dataset.

Table 2. IoT Train and Test Dataset Statistics

| IoT Device Train and Test Dataset Statistics | | | |
|---|---|---|---|
| Device | Rows | Columns | Size |
| Fridge | 59945 | 6 | 2,617KB |
| Garage Door | 59588 | 6 | 2,740KB |
| GPS Tracker | 58961 | 6 | 3,422KB |
| Modbus | 51107 | 8 | 2,959KB |
| Motion Light | 59489 | 6 | 2,416KB |

| Thermostat | 52775 | 6 | 2,547KB |
|---|---|---|---|
| Weather Sensor | 59261 | 7 | 4,132KB |

Table 3. IoT Train and Test Fridge Feature Descriptions

| IoT Train and Test Fridge Feature Descriptions | | | |
|---|---|---|---|
| ID | Feature | Type | Description |
| 1 | date | Date | Date of IoT logging |
| 2 | time | Time | Time of logging IoT data |
| 3 | fridge temperature | Number | Temperature measurement |
| 4 | temp_condition | String | Temperature conditions of the sensor, state whether it is high or low |
| 5 | label | Number | Normal or anomaly tag |
| 6 | Type | String | States the different types of attacks such as dos or backdoor. |

Table 4. IoT Train and Test Fridge Anomaly Statistics

| IoT Train and Test Fridge Anomaly Statistics | |
|---|---|
| Number of Rows | Type |
| 35000 | normal |
| 5000 | password |
| 2042 | XSS |
| 5000 | DDoS |
| 2902 | ransomware |
| 5000 | injection |
| 5000 | backdoor |

Table 5. IoT Train and Test Garage Door Feature Descriptions

| IoT Train and Test Garage Door Feature Descriptions | | | |
|---|---|---|---|
| ID | Feature | Type | Description |
| 1 | date | Date | Date of IoT logging |
| 2 | time | Time | Time of logging IoT data |
| 3 | Door_state | Boolean | State of the door sensor (true or false) |
| 4 | Sphone_signal | Boolean | State of the receiver of the door signal (true or false) |
| 5 | label | Number | Normal or anomaly tag |
| 6 | Type | String | States the different types of attacks such as dos or backdoor. |

Table 6. IoT Train and Test Garage Door Anomaly Statistics

| IoT Train and Test Garage Door Anomaly Statistics ||
|---|---|
| Number of Rows | Type |
| 35000 | normal |
| 5000 | password |
| 1156 | XSS |
| 5000 | DDoS |
| 2902 | ransomware |
| 5000 | injection |
| 5000 | backdoor |
| 529 | scanning |

Table 7. IoT Train and Test GPS profile Descriptions

| IoT Train and Test GPS Feature Descriptions ||||
|---|---|---|---|
| ID | Feature | Type | Description |
| 1 | date | Date | Date of IoT logging |
| 2 | time | Time | Time of logging IoT data |
| 3 | Latitude | Number | Latitude value of GPS tracker |
| 4 | Longitude | Number | Longitude value of GPS tracker |
| 5 | label | Number | Normal or anomaly tag |
| 6 | Type | String | States the different types of attacks such as dos or backdoor. |

Table 8. IoT Train and Test GPS Anomaly Statistic

| IoT Train and Test Anomaly Statistics ||
|---|---|
| Number of Rows | Type |
| 35000 | normal |
| 5000 | password |
| 577 | XSS |
| 5000 | DDoS |
| 2833 | ransomware |
| 5000 | injection |
| 5000 | backdoor |
| 550 | scanning |

Table 9. IoT Train and Test Modbus Feature Descriptions

| IoT Train and Test Modbus Feature Descriptions | | | |
|---|---|---|---|
| ID | Feature | Type | Description |
| 1 | date | Date | Date of IoT logging |
| 2 | time | Time | Time of logging IoT data |
| 3 | FC1_Read_Input_Register | Number | Modbus function that reads the input register |
| 4 | FC2_Read_Discrete_Value | Number | Modbus function that reads the discrete value |
| 5 | FC3_Read_Holding_Register | Number | Modbus function that reads the holding register |
| 6 | FC4_Read_Coil | Number | Modbus function that reads a coil |
| 7 | Label | Number | Normal or anomaly tag |
| 8 | Type | String | Different types of attacks such as dos or backdoor attacks. |

Table 10. IoT Train and Test Modbus Anomaly Statistics

| IoT Train and Test Modbus Anomaly Statistics | |
|---|---|
| Number of Rows | Type |
| 35000 | normal |
| 5000 | password |
| 577 | XSS |
| 5000 | injection |
| 5000 | backdoor |
| 529 | scanning |

Table 11. IoT Train and Test Motion Light Feature Description

| IoT Train and Test Motion Light Feature Descriptions | | | |
|---|---|---|---|
| ID | Feature | Type | Description |
| 1 | date | Date | Date of IoT logging |
| 2 | time | Time | Time of logging IoT data |
| 3 | motion_status | number | Status of motion light (0 or 1) |
| 4 | light_status | Boolean | Status of the light sensor (on or off) |
| 5 | Label | Number | Normal or anomaly tag |
| 6 | Type | String | Specifies types of attacks |

Table 12. IoT Train and Test Motion Light Anomaly Statistics

| IoT Train and Test Motion Light Anomaly Statistics ||
|---|---|
| Number of Rows | Type |
| 35000 | normal |
| 5000 | password |
| 449 | XSS |
| 5000 | DDoS |
| 2264 | ransomware |
| 5000 | injection |
| 5000 | backdoor |
| 1775 | scanning |

Table 13. IoT Train and Test Thermostat Feature Descriptions

| IoT Train and Test Thermostat Feature Descriptions ||||
|---|---|---|---|
| ID | Feature | Type | Description |
| 1 | date | Date | Date of IoT logging |
| 2 | time | Time | Time of logging IoT data |
| 3 | current_temp | Number | Temperature reading |
| 4 | thermostat_status | Boolean | Thermostat status (on or off) |
| 5 | Label | Number | Normal or anomaly tag |
| 6 | Type | String | Different types of attacks such as dos or backdoor attacks. |

Table 14. IoT Train and Test Thermostat Anomaly Statistics

| IoT Train and Test Thermostat Anomaly Statistics ||
|---|---|
| Number of Rows | Type |
| 35000 | normal |
| 5000 | password |
| 449 | XSS |
| 2264 | ransomware |
| 5000 | injection |
| 5000 | backdoor |
| 61 | scanning |

26Table 15. IoT Train and Test Weather Feature Descriptions

| IoT Train and Test Weather Feature Descriptions | | | |
|---|---|---|---|
| ID | Feature | Type | Description |
| 1 | date | Date | Date of IoT logging |
| 2 | time | Time | Time of logging IoT data |
| 3 | temperature | Number | Temperature measurement from sensor |
| 4 | pressure | Number | Pressure measurement from sensor |
| 5 | humidity | Number | Humidity measurement from sensor |
| 6 | Label | Number | Normal or anomaly tag |
| 7 | Type | String | Different types of attacks such as dos or backdoor attacks. |

Table 16. IoT Train and Test Weather Anomaly Statistics

| IoT Train and Test Weather Anomaly Statistics | |
|---|---|
| Number of Rows | Type |
| 35000 | normal |
| 5000 | password |
| 866 | xss |
| 5000 | ddos |
| 2865 | ransomware |
| 5000 | injection |
| 5000 | backdoor |
| 529 | scanning |

### 4.3 Processed Dataset

The ToN_IoT processed datasets were used to test whether the KMANB algorithm could be used on large pieces of data. The results were then examined to see whether the algorithm maintained its accuracy. These ToN_IoT processed datasets include more rows, file sizes and different amounts of anomalies. Table 17 shows the different devices, as well as the rows, columns, and size of each file. Table 18 to 24 depicts the number of anomalies present within each dataset.

Table 17. IoT Processed Dataset Statistics

| IoT Processed Dataset Statistics | | | |
|---|---|---|---|
| Device | Rows | Columns | Size |
| Fridge | 293009 | 6 | 13,116KB |
| Garage Door | 89754 | 6 | 4,286KB |
| GPS Tracker | 222325 | 6 | 30,010KB |
| Modbus | 198173 | 8 | 12,013KB |
| Motion Light | 242526 | 6 | 10,251KB |
| Thermostat | 185840 | 6 | 8,064KB |
| Weather Sensor | 251045 | 7 | 16,331KB |

Table 18. IoT Processed Fridge Anomaly Types

| IoT Processed Fridge Anomaly Types | |
|---|---|
| Number of Rows | Type |
| 206758 | normal |
| 28425 | password |
| 2042 | XSS |
| 10233 | DDoS |
| 2902 | ransomware |
| 7079 | injection |
| 35568 | backdoor |

Table 19. IoT Processed GPS Tracker Anomaly Types

| IoT Processed Garage Door Anomaly Types | |
|---|---|
| Number of Rows | Type |
| 25807 | normal |
| 16617 | password |
| 10230 | DDoS |
| 6331 | injection |
| 30230 | backdoor |
| 529 | scanning |

Table 20. IoT Processed GPS Anomaly Types

| IoT Processed GPS Tracker Anomaly Types | |
|---|---|
| Number of Rows | Type |
| 140488 | normal |
| 25176 | password |
| 577 | XSS |
| 10226 | DDoS |
| 2833 | ransomware |
| 6904 | injection |

| 35571 | backdoor |
|---|---|
| 550 | scanning |

Table 21. IoT Processed Modbus Anomaly Types

| IoT Processed Modbus Anomaly Types ||
|---|---|
| Number of Rows | Type |
| 133839 | normal |
| 18815 | password |
| 498 | XSS |
| 5186 | injection |
| 40005 | backdoor |
| 529 | scanning |

Table 22. IoT Processed Motion Light Anomaly Types

| IoT Processed Motion Light Anomaly Types ||
|---|---|
| Number of Rows | Type |
| 178591 | normal |
| 17521 | password |
| 449 | XSS |
| 8121 | DDoS |
| 2264 | ransomware |
| 5595 | injection |
| 28209 | backdoor |
| 1775 | scanning |

Table 23. IoT Processed Thermostat Anomaly Types

| IoT Processed Thermostat Anomaly Types ||
|---|---|
| Number of Rows | Type |
| 129563 | normal |
| 8435 | password |
| 449 | XSS |
| 2264 | ransomware |
| 9498 | injection |
| 35568 | backdoor |
| 61 | scanning |

Table 24. IoT Processed Weather Anomaly Types

| IoT Processed Weather Anomaly Types ||
|---|---|
| Number of Rows | Type |
| 160529 | normal |
| 25715 | password |
| 866 | XSS |
| 15182 | DDoS |
| 2865 | ransomware |
| 9726 | injection |
| 35641 | backdoor |
| 529 | scanning |

Both the train and test and the processed datasets were evaluated using WEKA (Waikato environment for knowledge analysis) version 3.8.4. This was done on a virtual machine within a Cyber Range, provided by Griffith University, School of Information and Communication Technology. WEKA was chosen because of its ease of use and its ability to utilise different algorithms.

## 4.4 Hypothesis

The hypothesis for this experiment is that the combination of UML and SML, forming HML will be of comparable strength to the traditional IoT ML algorithms. KMANB will also have similar if not quicker training and testing time than the more traditional SML algorithms. Furthermore, it is hypothesised that the KMANB will be more flexible compared to the other SML algorithms as well as being scalable in nature. The algorithm will be assessed by looking at the accuracy, precision and recall scores (APR scores), as well the training and testing times (speed). Although the ideas of flexibility and scalability are not easily quantified, these requirements will be assessed by looking at the overall APR scores over different devices and the larger dataset APR scores respectively. The APR scores are defined below:

## 4.5 Evaluation Metrics

WEKA's experiment function allows for the generation of true positive (TP), true negative (TN), false positive (FP) and false-negative (FN) numbers, which are then used to calculate the APR scores. TP refers to the number of correctly predicted packets, TN packets are predicted false and turn out to be false. FP are flagged as positive, with their true value being negative, and FN are predicted negatives, that are in fact true. The accuracy was calculated by adding the total number of true positives and true negatives together and then dividing them by the total of the true positives, true negatives, false negatives and the false positives. Precision involves true positives divided by the sum of the true positives and false positives. The recall involves dividing true positives by the sum of true positives and false negatives. These equations can be seen below (Javed et al., 2020). Furthermore, the training and testing times are also described below.

$$Accuracy = \frac{TP + TN}{TP + TN + FN + FP}$$

$$Precision = \frac{TP}{TP + FP}$$

$$Recall = \frac{TP}{TP + FN}$$

After the experiments are run, the number of TP, TN, FP, FN is presented by WEKA. These will then be placed into a spreadsheet with the accuracy (Acc), precision (Pre) and recall (Rec) formulas. This will then present the calculated APR scores. It should also be noted that CFS was used to evaluate both the train and test datasets as well as the larger processed datasets. This was done in order to find the highest ranked correlation to the type of anomaly, and in aid in its removal to examine the feature reduction research question.

The suitability for these metrics refers to the idea of what makes a strong AD system. A strong AD system would need something accurate, precise and can be reapplied.

## 5 Results

The results of the KMANB HML algorithm will be compared to 3 other SML algorithms using the Train and Test dataset as a baseline. This will be done to demonstrate proof of concept and to examine the performance of our algorithm compared to traditional SML methods.

### 5.1 Train and Test Results

The below tables in this section are involve three SML algorithms being compared against the proposed KMANB algorithm.

Table 25. IoT Train and Test Fridge Experiment Results

| IoT Train and Test Fridge Experiment Results | | | | |
|---|---|---|---|---|
| | Random Forest | Naive Bayes | KNN | KMANB |
| Accuracy | 0.97 | 0.53 | 0.99 | 0.99 |
| Precision | 0.97 | 0.53 | 0.99 | 0.99 |
| Recall | 0.97 | 0.51 | 0.99 | 0.99 |
| Train Time | 0.188 | 0.011 | 0.147 | 2.98 |
| Test Time | 0.045 | 0.005 | 2.556 | 0.44 |

Table 26. IoT Train and Test Garage Door Experiment Results

| IoT Train and Test Garage Door Experiment Results | | | | |
|---|---|---|---|---|
| | Random Forest | Naive Bayes | KNN | KMANB |
| Accuracy | 1 | 1 | 1 | 1 |
| Precision | 1 | 1 | 1 | 1 |
| Recall | 1 | 1 | 1 | 1 |
| Train Time | 0.062 | 0.010 | 0.625 | 3.38 |
| Test Time | 0.000 | 0.002 | 0.969 | 0.48 |

Table 27.  IoT Train and Test GPS Tracker Experiment Results

| IoT Train and Test GPS Tracker Experiment Results | | | | |
|---|---|---|---|---|
| | Random Forest | Naive Bayes | KNN | KMANB |
| Accuracy | 0.85 | 0.84 | 0.88 | 0.99 |
| Precision | 0.85 | 0.86 | 0.89 | 0.99 |
| Recall | 0.85 | 0.85 | 0.88 | 0.95 |
| Train Time | 0.833 | 0.009 | 0.08 | 5.39 |
| Test Time | 0.099 | 0.007 | 1.508 | 0.78 |

Table 28. IoT Train and Test Modbus Experiment Results

| IoT Train and Test Modbus Experiment Results | | | | |
|---|---|---|---|---|
| | Random Forest | Naive Bayes | KNN | KMANB |
| Accuracy | 0.97 | 0.67 | 0.77 | 0.98 |
| Precision | 0.98 | 0.46 | 0.77 | 0.95 |
| Recall | 0.98 | 0.68 | 0.78 | 0.96 |
| Train Time | 1.587 | 0.012 | 0.060 | 4.97 |
| Test Time | 0.031 | 0.002 | 0.116 | 0.70 |

Table 29. IoT Train and Test Motion Light Experiment Results

| IoT Train and Test Motion Light Experiment Results | | | | |
|---|---|---|---|---|
| | Random Forest | Naive Bayes | KNN | KMANB |
| Accuracy | 0.58 | 0.58 | 0.54 | 1 |
| Precision | 0.34 | 0.34 | 0.34 | 1 |
| Recall | 0.59 | 0.59 | 0.59 | 1 |
| Train Time | 0.1 | 0.011 | 3.157 | 2.44 |
| Test Time | 0.008 | 0.002 | 6.409 | 0.34 |

Table 30. IoT Train and Test Thermostat Experiment Results

| IoT Train and Test Thermostat Experiment Results | | | | |
|---|---|---|---|---|
| | Random Forest | Naive Bayes | KNN | KMANB |
| Accuracy | 0.66 | 0.66 | 0.60 | 0.99 |
| Precision | 0.55 | 0.44 | 0.56 | 0.97 |
| Recall | 0.66 | 0.66 | 0.61 | 0.93 |
| Train Time | 1.044 | 0.009 | 0.064 | 4.13 |
| Test Time | 0.023 | 0.002 | 0.088 | 0.62 |

Table 31. IoT Train and Test Weather Experiment Results

| . IoT Train and Test Weather Experiment Results | | | | |
|---|---|---|---|---|
| | Random Forest | Naive Bayes | KNN | KMANB |
| Accuracy | 0.84 | 0.69 | 0.81 | 0.97 |

3232

| Precision | 0.84 | 0.72 | 0.81 | 0.87 |
|---|---|---|---|---|
| Recall | 0.84 | 0.69 | 0.81 | 0.90 |
| Train Time | 0.789 | 0.011 | 0.066 | 6.28 |
| Test Time | 0.008 | 0.002 | 0.414 | 0.92 |

## 5.2 Train and Test with No Highest Ranked Feature

The below results are the Train and Test datasets with the feature Date removed. CFS was run in order to find the feature with the highest correlation to anomaly type and remove it. This was done in order to compare the training and testing times and investigate whether data reduction decreases the train and test time, subsequently negatively impacting the APR scores. If this algorithm produces strong results even with the removal of the highest-ranked correlation, it will further solidify the idea that it is a reasonable alternative to the traditional SML algorithms.

It should also be noted that KMANB is the only algorithm being tested to see if feature reduction improves the training and testing times whilst maintaining the APR scores. As such, it was not necessary to include the traditional SML algorithms times in the below tables.

Table 32. IoT Train and Test Fridge with No Highest Ranked Experiment Results

| IoT Train and Test Fridge with No Highest Ranked Experiment Results | | | | |
|---|---|---|---|---|
| | Random Forest | Naive Bayes | KNN | KMANB |
| Accuracy | 0.97 | 0.53 | 0.99 | 0.99 |
| Precision | 0.97 | 0.53 | 0.99 | 0.99 |
| Recall | 0.97 | 0.51 | 0.99 | 0.99 |
| Training Time | | | | 2.88 |
| Testing Time | | | | 0.44 |

Table 33. IoT Train and Test Garage Door with No Highest Ranked Experiment Results

| IoT Train and Test Garage Door with No Highest Ranked Experiment Results | | | | |
|---|---|---|---|---|
| | Random Forest | Naive Bayes | KNN | KMANB |
| Accuracy | 1 | 1 | 1 | 0.99 |
| Precision | 1 | 1 | 1 | 0.99 |
| Recall | 1 | 1 | 1 | 0.99 |
| Training Time | | | | 0.92 |
| Testing Time | | | | 0.12 |

Table 34. IoT Train and Test GPS Tracker with No Highest Ranked Experiment Results

| IoT Train and Test GPS Tracker with No Highest Ranked Experiment Results | | | | |
|---|---|---|---|---|
| | Random Forest | Naive Bayes | KNN | KMANB |
| Accuracy | 0.85 | 0.84 | 0.88 | 0.99 |
| Precision | 0.85 | 0.86 | 0.89 | 0.94 |
| Recall | 0.85 | 0.85 | 0.88 | 0.96 |
| Training Time | | | | 4.39 |
| Testing Time | | | | 0.66 |

Table 35. IoT Train and Test Modbus with No Highest Ranked Experiment Results

| IoT Train and Test Modbus with No Highest Ranked Experiment Results | | | | |
|---|---|---|---|---|
| | Random Forest | Naive Bayes | KNN | KMANB |
| Accuracy | 0.97 | 0.67 | 0.77 | 0.98 |
| Precision | 0.98 | 0.46 | 0.77 | 0.96 |
| Recall | 0.98 | 0.68 | 0.78 | 0.95 |
| Training Time | | | | 4.43 |
| Testing Time | | | | 0.64 |

Table 36. IoT Train and Test Motion Light with No Highest Ranked Experiment Results

| IoT Train and Test Motion Light with No Highest Ranked Experiment Results | | | | |
|---|---|---|---|---|
| | Random Forest | Naive Bayes | KNN | KMANB |
| Accuracy | 0.58 | 0.58 | 0.54 | 0.99 |
| Precision | 0.34 | 0.34 | 0.34 | 0.99 |
| Recall | 0.59 | 0.59 | 0.59 | 1 |
| Training Time | | | | 1.63 |
| Testing Time | | | | 0.22 |

Table 37. IoT Train and Test Thermostat with No Highest Ranked Experiment Results

| IoT Train and Test Thermostat with No Highest Ranked Experiment Results | | | | |
|---|---|---|---|---|
| | Random Forest | Naive Bayes | KNN | KMANB |
| Accuracy | 0.66 | 0.66 | 0.60 | 0.99 |
| Precision | 0.55 | 0.44 | 0.56 | 0.97 |
| Recall | 0.66 | 0.66 | 0.61 | 0.91 |
| Training Time | | | | 3.49 |
| Testing Time | | | | 0.54 |

Table 38. IoT Train and Test Weather with No Highest Ranked Experiment Results

| IoT Train and Test Weather with No Highest Ranked Experiment Results | | | | |
|---|---|---|---|---|
| | Random Forest | Naive Bayes | KNN | KMANB |
| Accuracy | 0.84 | 0.69 | 0.81 | 0.97 |
| Precision | 0.84 | 0.72 | 0.81 | 0.89 |
| Recall | 0.84 | 0.69 | 0.81 | 0.91 |
| Training Time | | | | 5.24 |
| Testing Time | | | | 0.79 |

### 5.3 Processed Dataset Results

The train and test datasets were used as a proof of concept. The processed datasets results are examined to test the scalability of the KMANB, and to ascertain if it could be applicable in real world situations on larger datasets.

Table 39. IoT Processed Dataset Experiment Results

| | Fridge | Garage Door | GPS Tracker | Modbus | Motion Light | Thermostat | Weather |
|---|---|---|---|---|---|---|---|
| IoT Processed Dataset Experiment Results ||||||||
| KMANB Results ||||||||
| Accuracy | 0.99 | 0.99 | 0.98 | 0.98 | 0.99 | 0.98 | 0.98 |
| Precision | 0.99 | 0.99 | 0.92 | 0.96 | 0.99 | 0.95 | 0.89 |
| Recall | 0.99 | 0.99 | 0.82 | 0.95 | 0.98 | 0.95 | 0.92 |

### 5.4 Processed Dataset No Highest Ranked Feature

The processed dataset with the no highest ranked results are recorded below. This was done to ascertain whether the highest ranked correlation to the anomaly type, would affect the APR results in the larger datasets, the same as the above smaller datasets.

Table 40. IoT Processed Dataset with No Highest Ranked Experiment Results

| | Fridge | Garage Door | GPS Tracker | Modbus | Motion Light | Thermostat | Weather |
|---|---|---|---|---|---|---|---|
| IoT Processed Dataset with No Highest Ranked Experiment Results ||||||||
| KMANB Results ||||||||
| Accuracy | 0.99 | 0.99 | 0.99 | 0.97 | 0.99 | 0.99 | 0.98 |
| Precision | 0.97 | 0.99 | 0.92 | 0.89 | 0.99 | 0.98 | 0.93 |
| Recall | 0.98 | 1 | 0.94 | 0.93 | 0.99 | 0.98 | 0.94 |

## 6 Discussion/Analysis

The results from the KMANB algorithm experiments will be described looking at the following key criteria: accuracy, speed, scalability and flexibility, as well as taking into consideration the associated APR scores. The experiments will also answer the research questions posed by this thesis.

### 6.1 Does the proposed algorithm have higher accuracy, precision and recall scores than traditional SML?

Firstly, it can be suggested that the KMANB is overall stronger regarding accuracy, precision and recall within the different IoT systems.

Looking at the first section with the smaller, train and test datasets, the proposed algorithm scored higher APR scores than RF, NB and KNN. This was seen as table 25, shows KMANB scored with a 0.99 on all 3 of APR, as opposed to the RF algorithm, which scored a 0.97 on all 3 measures, and the NB which scored 0.53 on the accuracy, precision, and a 0.51 on the recall aspect. Table 26 showed that all tested algorithms gave APR scores as 1, for all. The Table 27 experiments found that once again, the KMANB was the highest rated across all three APR measurements, with the scores being presented at 0.99, 0.99, and 0.95 respectively. This is compared to the RF scores of 0.85, 0.85, 0.85, NB with 0.84, 0.86, 0.85, and KNN scores of 0.88, 0.89, 088 respectively. Table 28 results demonstrated that overall, the KMANB was accurate to a certain extent. This was evident as the accuracy measurement was 0.98 versus 0.77 for KNN, 0.67 for NB and RF measuring at 0.97. However, the RF scored 0.98 for precision and recall, as opposed to 0.95 and 0.96 for the KMANB. KMANB scored second-highest overall, as compared to NB and KNN but still less than the standard RF algorithm. Table 29 showed that KMANB is accurate for this type of IoT device. The APR score was 1,1 and 1 respectively. Compared to the RF and its rating of 0.58, 0.34, and 0.59, NB with 0.66, 0.44 and 0.66 as well as KNN with 0.60, 0.56, and 0.61. Table 30 shows ratings of 0.99, 0.97 and 0.93 respectively for the proposed algorithm. This contrasts with RF and its 0.66, 0.55 and 0.66, NB and 0.66, 0.44 and 0.66, and lastly KNN and 0.60, 0.56 and 0.61. Table 31 also reinforced the idea of the KMANB's accuracy, with ratings of 0.99, 0.97 and 0.93 respectively. Once again, this is compared to the scores for RF (0.66, 0.55 and 0.66), NB (0.66, 0.44 and 0.66) and KNN (0.60, 0.56 and 0.61). There were several interesting results in terms of APR that have been found. One result was that Table 26's scores were all 1. This could be because of several different reasons, although Alsaedi et al. (2020) suggests that it's because of the type of data that is found within the dataset. The data is of a discrete nature, meaning that it only has a certain amount of values to be counted, an example of this is the number of students in a classroom. This means that each algorithm might have been able to adequately predict each outcome as the data.

These above scores demonstrate the overall strength of the KMANB, and show that compared to traditional SML algorithms, it is just as, if not a stronger choice. The accuracy of the KMANB could be due to the pre-classification step of the KM algorithm. As the KM algorithm was used to generate a new cluster within the dataset, the NB would have theoretically seen that cluster, ingested and trained from it. After the training took place, the testing step would have then utilised what was learnt from the already clustered classes within the dataset, and then based its predictions from that. This is also reinforced by a strength of HML, as UML could theoretically help SML with providing more data points to base its labels from.

### 6.2 Does the proposed algorithm have faster train and test times than the traditional SML algorithms?

The speed (train and test time) for the KMANB was not as competitive as first thought. Looking at Table 25, the KMANB test came in at 2.98 seconds and then 0.44 for a total time of operation for 3.42 seconds. This is far behind the time of RF, NB and KNN with total times of 0.233, 0.115 and 2.703 respectively. The scores for fig 26 are logged as 3.38 train and 0.48 test time, meaning a total time of 3.86 was recorded. This once again in comparison to the scores of the RF, NB and KNN of 0.062, 0.012 and 1.594 respectively. The Table 27 KMANB speed times are a total of 6.17 total time (5.39 train and 0.78 test). This is contrasted to the scores of 0.932, 0.016 and 0.08 for the 3 SML algorithms. Table 28 results suggested that once again, our proposed algorithm operated several seconds behind, with a training time of 4.97 and a testing time of 0.70 giving a total operation time of 5.67. The scores for the RF, NB and KNN were found to be in totality, 1.618, 0.014 and 0.176. The Motion light results showed that KMANB was slower with a total time of operation for 2.78. The RF and NB times were found to be quicker, with total times of 0.108, 0.01. However, it should be noted that the KNN time was far slower with a total operation time of 9.566, compared to the proposed algorithm. Table 30 found the KMANB was slower once again. The RF, NB and KNN algorithms were found to be run at 1.067, 0.011 and 0.152 respectively while the KMANB spent a total of 4.75 seconds in operation. Lastly, Table 31 showed that the KMANB spent 7.2 seconds in operation, versus the RF, NB and KNN scores of 0.797, 0.013 and 0.48 seconds of operation.

It should be noted that Table 29's scores for KMANB were slower than the RF and NB scores. This could be for several reasons. As KNN assumes that everything that is close, is related, it would try to classify data that is already grouped together, however light_status as shown in Table 11, is Boolean. Meaning its either True or False (on or off), as such this might have affected the accuracy of class membership designation. Furthermore, the way in which KNN operates, requires the total ingestion of the data before calculations can occur. This means that the algorithm would have had to essentially load everything up first. This is contrasted to NB that only works on assumptions of independence, not requiring the entire dataset at once to be ingested.

The slower times of the KMANB could be affected by the hardware the test was run on as this can either speed up or slow down the operation. As our tests were run on a VM with 2 CPU's ~2.7GHZ, our scores could have been negatively impacted by this. However, this is not suggesting that KMANB is quicker, it does in fact appear to be generally slower, but just not as slow as is being shown. These results suggest that the proposed algorithm is not as fast comparatively to the traditional SML algorithms. This leads to the idea of examining what is the correct amount of data reduction needed to find the balance between speed and precision of an AD algorithm. Too little accuracy and the algorithm is useless, too little speed and it becomes redundant. The following No Highest Ranked Feature dataset experiments were run to ascertain if our algorithm's speed could be improved whilst maintaining its APR scores.

### 6.3   Does removing the highest ranked subset of attributes to the type of anomaly, negatively impact the strength or speed of the algorithm?

Firstly, looking at Table 32, the training and testing times of the KMANB were still high comparatively, coming in at 2.88 and 0.44. Table 33 yielded 0.92 training and 0.12 testing times. The Table 34 training set gave 4.39 training and 0.66 testing time. The Table 35 training and testing times produced 4.43 and 0.64 times respectively. Table 36 being found at a 1.63 training and a 0.22 testing time. Table 37 was found to have times of 3.49 and 0.54 respectively. Lastly, Table 38 registered a time of 5.24 and a 0.79.

If Table 32 is examined further, KMANB is ranked at 2.88 and 0.44 train and test speed, compared to Fig 25 which was 2.98 and 0.44. This was due in large to the highest correlation feature of date being removed and it not affecting the score. This means that the reduction of data leads to KMANB being able to operate quicker. Another Table that can be examined, would be the Weather Train and Test. Without the highest ranked feature correlation, its scores are 5.24 and 0.79 for speed, as well as 0.97, 0.89 and 0.91 for APR, respectively. Compared to Table 31, and the scores of 6.28 and 0.92 for speed, and 0.97, 0.87 and 0.90 APR scores, respectively. The drop in the speed and the increase in the precision and recall for Table 38 could be attributed to the highest feature correlation having too much weight and therefore influencing the algorithm and obfuscating some of the rankings. This could be because date is nominal and one nominal attribute is counted as 1 in WEKA, if there are 100,000 packets on 1 day, that is 100,000 added to the weight of that day. This means that more numbers and a greater distribution of weighting is allocated to this feature. When date is removed, memory is freed up and the subsequent weight given to this attribute is removed.

Overall, the results suggest that accuracy was not sacrificed to the point where the algorithm becomes unreliable. An example of this is the garage door IoT (Table 33) dropping by 0.01 to 0.99 for all the APR measurements, whilst the train and test speed improving from 3.68 and 0.18, training and testing to 2.88 and 0.44. It should be noted that overall, the total time in operation is lowered by 0.54 seconds. The testing time increases marginally, this is due in large to the algorithm having less data to base its predictions off, but enough that it is not thrown out considerably. The trade-off for speed and accuracy stems from the idea that feature reduction is needed to ensure a faster, more precise algorithm when regarding anomaly detection within networks. It could be suggested that this is not specifically needed for IoT sensor data, as it is already less complicated in nature compared to the traditional TCP/IP data found within networking. Another result which reinforces this involves Table 38. The no highest results demonstrated that although the accuracy was found to be the same, the precision was slightly more with a 0.02 increase but a decrease of 0.01 regarding the recall score. The time in training was found to be a total of 1.17 seconds slower. This demonstrates that within IoT sensor data, the trade-off can be achieved with minimal downside, further improving the algorithm and its chances of adoption however, it is not mandatory. Arguably, the most important facet of an AD algorithm is accuracy, and the ability to consistently produce strong results.

It should also be noted that the original plans for this algorithm involved adding an additional layer of KM clustering to go to 2 total clusters, clustered to normal and anomaly. The algorithm would then be further clustered to the formula of $C=A+1$. This was done to see if adding an additional cluster assist with the final step and to produce higher results. It was found that it did not help, and as such it was decided not to be included within the final algorithm. This once again forced the examination of the trade-off between speed and accuracy. It was assessed as adding another layer of algorithm, increasing complexity, with no real gain.

### 6.4 Does the proposed algorithm maintain its strength as it operates on larger datasets?

The scalability of the KMABNB is also being examined. Scalability refers to the idea that once this has been applied to the smaller, even datasets, this concept is then taken and applied to larger, uneven datasets. Table 39 depicts all the results for each type of IoT. The fridge scores 0.99 for all 3 types of ratings, as does the garage door. The GPS tracker rates at 0.98, 0.92 and 0.82 for APR respectively. The Modbus results indicate a 0.98, 0.96 and 0.95 APR score. The motion light was found to have results of 0.99 for accuracy and precision, with a recall ranking of 0.98. The KMANB on thermostat was given an APR score of 0.98, 0.95 and 0.95, with the weather dataset's run resulting in a 0.98, 0.89 and 0.92 in scores. Looking at this, it can be suggested this algorithm is indeed scalable and even resulted in comparable readings in the bigger datasets, when compared to the smaller datasets. An example of this is the fridge's APR results were 0.99 for all the areas, the same as the larger datasets. Furthermore, the garage door was slightly less on the processed side, with 0.99 for all 3, compared to the smaller datasets with full 1's. Although this is indeed less, it still doesn't dissuade from the notion that this algorithm is indeed scalable, as a loss of 0.01 for all 3 sections still points to overall a strong algorithm. If the smaller weather dataset's results are examined, this one does indicate the larger dataset produced stronger results. The APR results for the smaller dataset were 0.97, 0.87 and 0.90, compared to 0.98, 0.89 and 0.92 respectively. This further demonstrates the scalable nature of this algorithm, as it is even, if not higher rated for the use of IoT anomaly detection with both smaller and larger datasets. Table 40's results also reiterated the trade-off between accuracy and speed. It depicts the APR scores of the fridge, garage door, GPS tracker, Modbus, motion light, thermostat and weather sensor. It shows that KMANB still maintained strength even with the removal of the highest correlation to anomaly type.

### 6.5 How much better would a hybrid machine learning algorithm comprised, of KM clustering and Naive Bayes be than traditional SML algorithms when it comes to AD within IoT sensor data?

The KMANB algorithm appears to be a stronger choice for all of the IoT devices listed above. The main reason why due to its higher APR scores, which indicates a stronger algorithm when it comes to AD. Furthermore, it appears to be flexible, maintaining its high scores across multiple devices and dataset sizes. This means that the research question "Is the proposed algorithm able to be applied to different types of IoT devices?" Is sufficiently answered. Although the KMANB is slower, it consistently provides the same APR scores across different devices, both including the smaller and larger datasets. Thus, the requirement of scalability is met.

## 7  Conclusion

The research questions found within this document seek to discuss several facets of the proposed KMANB algorithm and its use on IoT sensor data. The first of these questions is arguably the most important one, and that is whether the KMANB is the stronger choice compared to traditional SML algorithms. Once this question has been answered, this then allows the further examination of sub theories. One sub theory relates to whether the KMANB is faster in terms of total time in operation compared to the other SML algorithms. This was done by first utilising the train and test IoT sensor datasets and comparing the APR results for our data, against traditional SML algorithms. Next, the scalability of the proposed algorithm was discussed as well, with the KMANB needing to be used to test large slices of data to be found viable. This was achieved by using the processed datasets and seeing if our proposed algorithm was still accurate. Finally, the trade-off between speed and accuracy of the KMANB was also tested. This was done by comparing the two smaller datasets with one having the largest feature correlation relating the type of anomaly removed. The speed and APR measures were then compared to ascertain whether or not the score reductions were drastic and algorithm breaking. The research questions were all met, as our proposed algorithm rated higher in almost all the APR tests overall, maintained its high scores on the larger, uneven datasets. However, it should be noted that its speed was slower in some instances. This speed issue was addressed also, with the reduction of the highest feature correlation of an anomaly occurring. This resulted in the lowering of the time in operation on some devices.

## 7.1 Future Work

This algorithm and its application to IoT sensor data has multiple avenues to be further examined. One such potential future work involves the isolation of the IoT anomalies. As we have successfully identified them, further research could be completed in which some form of Principal Component Analysis is carried out on the data that is found to be anomalous. This could be done by exporting the correctly predicted data into a .csv file, and then loaded into WEKA. After this has occurred, PCA could be run, and then this could potentially be able to show what sensor information is aligned to the correctly predicted anomalies. Other future work that could be carried out regards the increase in scope for the KMANB and its use. As mentioned above, there are millions of different IoT devices, and as such the scope of this could be further increased to test specific brands of garage doors, fridges or GPS trackers. This would theoretically allow us to further examine the potential real-world applications for the proposed algorithm. Alternate reduction of different data could also be further examined. As the highest correlation to the type of anomaly was reduced in the no highest ranked feature datasets, further reduction of the biggest in terms of size could be done, as this would potentially not reduce the accuracy by a considerable amount and also help to reduce the time in operation.

41Jagadish, H., Ooi, B., Tan, K., Yu, C., & Zhang, R. (2005). iDistance: An adaptive B + -tree based indexing method for nearest neighbor search. *ACM Transactions on Database Systems*, 30(2), 364-397. https://doi.org/10.1145/1071610.1071612

Kassab, W., & Darabkh, K. (2020). A-Z survey of internet of things: Architectures, protocols, applications, recent advances, future directions and recommendations. *Journal of Network and Computer Applications, 163*. doi:10.1016/j.jnca.2020.102663

Khamparia, A., Pande, S., Gupta, D., Khanna, A., & Sangaiah, A. (2020). Multi-level framework for anomaly detection in social networking. *Library Hi Tech, 38*, 350–366. doi:10.1108/LHT-01-2019-0023

Kristianto, R., Santoso, B., & Sari, M. (2019). Integration of K-means clustering and naïve bayes classification algorithms for smart AC monitoring and control in WSAN. *Integration of K-means clustering and naïve bayes classification algorithms for smart AC monitoring and control in WSAN*. doi:10.1109/ICITISEE48480.2019.9003927 65

Lawal, M., Shaikh, R., & Hassan, S. (2020). An anomaly mitigation framework for iot using fog computing. *Electronics (Basel), 9*, 1–24. doi:10.3390/electronics9101565

Li, N., Martin, A., & Estival, R. (2018). Combination of supervised learning and un-supervised learning based on object association for land cover classification. *Combination of supervised learning and unsupervised learning based on object association for land cover classification*. doi:10.1109/DICTA.2018.8615871

Memon, R., Li, J., Nazeer, M., Khan, A., & Ahmed, J. (2019). DualFog-IoT: Additional fog layer for solving blockchain integration problem in internet of things. *IEEE Access, 7*, 169073–169093. doi:10.1109/ACCESS.2019.2952472

Naik, N. (2017). Choice of effective messaging protocols for IoT systems: MQTT, CoAP, AMQP and HTTP. *Choice of effective messaging protocols for IoT systems: MQTT, CoAP, AMQP and HTTP*. doi:10.1109/SysEng.2017.8088251

Om, H., & Kundu, A. (2012). A hybrid system for reducing the false alarm rate of anomaly intrusion detection system. *A hybrid system for reducing the false alarm rate of anomaly intrusion detection system*. doi:10.1109/RAIT.2012.6194493

Patel, C., & Doshi, N. (2020). A novel MQTT security framework in generic IoT model. *Procedia Computer Science, 171*, 1399–1408. doi:10.1016/j.procs.2020.04.150

Qi, J., Yu, Y., Wang, L., Liu, J., & Wang, Y. (2017). An effective and efficient hierarchical K-means clustering algorithm. *International Journal of Distributed Sensor Networks, 13*, 1–17. doi:10.1177/1550147717728627

42Quek, Y., Woo, W., & Thillainathan, L. (2020). IoT load classification and anomaly warning in ELV DC picogrids using hierarchical extended -nearest neighbors. *IEEE Internet of Things Journal, 7*, 863–873. doi:10.1109/JIOT.2019.2945425 66

Radoglou Grammatikis, P. I., Sarigiannidis, P. G., & Moscholios, I. D. (2019). Securing the internet of things: Challenges, threats and solutions. *Internet of Things*, 5, 41-70. https://doi.org/10.1016/j.iot.2018.11.003

Rawat, D., & Reddy, S. (2017). Software defined networking architecture, security and energy efficiency: A survey. *IEEE Communications Surveys and Tutorials, 19*, 325–346. doi:10.1109/COMST.2016.2618874

Sahu, N., & Mukherjee, I. (2020). Machine learning based anomaly detection for IoT network: (anomaly detection in IoT network). *Machine learning based anomaly detection for IoT network: (anomaly detection in IoT network)*. doi:10.1109/ICOEI48184.2020.9142921

Samrin, R., & Vasumathi, D. (2018). Hybrid weighted K-means clustering and artificial neural network for an anomaly-based network intrusion detection system. *Journal of Intelligent Systems, 27*, 135–147. doi:10.1515/jisys-2016-0105

Saputra, M., Widiyaningtyas, T., & Wibawa, A. (2018). Illiteracy classification using K means-naïve bayes algorithm. *JOIV : International Journal on Informatics Visualization, 2*, 153–158. doi:10.30630/joiv.2.3.129

Sethi, P., & Sarangi, S. (2017). Internet of things: Architectures, protocols, and applications. *Journal of Electrical and Computer Engineering*, 1–25. doi:10.1155/2017/9324035

Sharma, S., Pandey, P., Tiwari, S., & Sisodia, M. (2012). An improved network intrusion detection technique based on k-means clustering via naïve bayes classification. *An improved network intrusion detection technique based on k-means clustering via naïve bayes classification*, 417–422.

Soe, Y., Feng, Y., Santosa, P., Hartanto, R., & Sakurai, K. (2020). *Machine learning-based IoT-botnet attack detection with sequential architecture. Sensors*, 4372. doi:10.3390/s20164372 67

Tayal, D., Jain, A., & Meena, K. (2016). Development of anti-spam technique using modified K-means & naive bayes algorithm. *Development of anti-spam technique using modified K-means & naive bayes algorithm*, 2593–2597.
Teh, H. Y., Wang, K. I., & Kempa-Liehr, A. W. (2021). *Expect the unexpected: Unsupervised feature selection for automated sensor anomaly detection. IEEE Sensors Journal,* , 1-1. https://doi.org/10.1109/JSEN.2021.3084970